\documentclass[a4paper,11pt]{article}

\usepackage{jinstpub} 
\usepackage{arydshln} 
\usepackage{amssymb}
\usepackage{amsmath}
\usepackage{threeparttable}
\usepackage{lineno}
\usepackage[normalem]{ulem}
\usepackage{lscape}
\usepackage{graphicx}

\begin{document}

\title{\boldmath Development of low-background photomultiplier tubes for liquid xenon detectors}

\author[a,e]{K.~Abe,}
\author[a]{Y.~Chen,}
\author[a,e]{K.~Hiraide,}
\author[a,e,1]{K.~Ichimura,}
\author[a]{S.~Imaizumi,}
\author[a]{N.~Kato,}
\author[a,1,e]{Y.~Kishimoto,\note{Now at Research Center for Neutrino Science, Tohoku University, Sendai 980-8578, Japan}}
\author[a,e,2]{K.~Kobayashi,\note{Now at Waseda Research Institute for Science and Engineering, Waseda University, Shinjuku, Tokyo 169-8555, Japan}}
\author[a,3]{M.~Kobayashi,\note{Now at Physics Department, Columbia University, New York, NY 10027, USA}}
\author[a,e]{S.~Moriyama,}
\author[a,e]{M.~Nakahata,}
\author[a,4]{K.~Sato,\note{Now at Institute for Space-Earth Environmental Research, Nagoya University, Nagoya, Aichi 464-8601, Japan}}
\author[a,e]{H.~Sekiya,}
\author[a]{T.~Suzuki,}
\author[a,e]{A.~Takeda,}
\author[a,5]{S.~Tasaka,\note{Now at Gifu University, Gifu 501-1193, Japan}}
\author[a,e,4]{M.~Yamashita,}
\author[b]{B.~S.~Yang,}
\author[c]{N.~Y.~Kim,}
\author[c]{Y.~D.~Kim,}
\author[c,h]{Y.~H.~Kim,}
\author[d]{R.~Ishii,}
\author[d,f]{Y.~Itow,}
\author[d]{K.~Kanzawa,}
\author[d]{K.~Masuda,}
\author[e]{K.~Martens,}
\author[e,6]{A.~Mason,\note{Now at Department of Physics, University of Oxford, Oxford, Oxfordshire, United Kingdom}}
\author[e,7]{Y.~Suzuki,\note{Now at Institute for Cosmic Ray Research, the University of Tokyo, Kashiwa, Chiba 277-8582, Japan}}
\author[g]{K.~Miuchi,}
\author[g,e]{Y.~Takeuchi,}
\author[h]{K.~B.~Lee,}
\author[h]{M.~K.~Lee,}
\author[i]{Y.~Fukuda,}
\author[j,e]{H.~Ogawa,}
\author[k]{K.~Nishijima,}
\author[l]{K.~Fushimi,}
\author[m,e]{B.~D.~Xu,}
\author[n]{S.~Nakamura}

\begin{center}
\collaboration[c]{XMASS Collaboration}
\end{center}

\affiliation[a]{Kamioka Observatory, Institute for Cosmic Ray Research, the University of Tokyo, Higashi-Mozumi, Kamioka, Hida, Gifu 506-1205, Japan}
\affiliation[b]{Center for Axion and Precision Physics Research, Institute for Basic Science, Daejeon 34051, South Korea }
\affiliation[c]{Center for Underground Physics, Institute for Basic Science, 70 Yuseong-daero 1689-gil, Yuseong-gu, Daejeon 305-811, South Korea}
\affiliation[d]{Institute for Space-Earth Environmental Research, Nagoya University, Nagoya, Aichi 464-8601, Japan}
\affiliation[e]{Kavli Institute for the Physics and Mathematics of the Universe (WPI), the University of Tokyo, Kashiwa, Chiba 277-8582, Japan}
\affiliation[f]{Kobayashi-Maskawa Institute for the Origin of Particles and the Universe, Nagoya University, Furo-cho, Chikusa-ku, Nagoya, Aichi 464-8602, Japan}
\affiliation[g]{Department of Physics, Kobe University, Kobe, Hyogo 657-8501, Japan}
\affiliation[h]{Korea Research Institute of Standards and Science, Daejeon 305-340, South Korea}
\affiliation[i]{Department of Physics, Miyagi University of Education, Sendai, Miyagi 980-0845, Japan}
\affiliation[j]{Department of Physics, College of Science and Technology, Nihon University, Kanda, Chiyoda-ku, Tokyo 101-8308, Japan}
\affiliation[k]{Department of Physics, Tokai University, Hiratsuka, Kanagawa 259-1292, Japan}
\affiliation[l]{Department of Physics, Tokushima University, 2-1 Minami Josanjimacho Tokushima city, Tokushima 770-8506, Japan}
\affiliation[m]{Department of Engineering Physics, Tsinghua University, Haidian District, Beijing, China 100084}
\affiliation[n]{Department of Physics, Faculty of Engineering, Yokohama National University, Yokohama, Kanagawa 240-8501, Japan}

\emailAdd{xmass.publications16@km.icrr.u-tokyo.ac.jp}

\abstract{
  We successfully developed a
  new photomultiplier tube (PMT) with a three-inch diameter, convex-shaped photocathode, R13111.
  Its prominent features include  good performance
  and ultra-low radioactivity.
  The convex-shaped photocathode realized a large photon acceptance
  and good timing resolution.   
  Low radioactivity was achieved by three factors:
  (1) the glass material was synthesized using low-radioactive-contamination material; 
  (2) the photocathode was produced with $^{39}$K-enriched potassium;
  and 
  (3) the purest grade of aluminum material was used for the vacuum seal.
  As a result each R13111 PMT contains only about 0.4 mBq of $^{226}$Ra,
  less than 2 mBq of $^{238}$U, 0.3 mBq of $^{228}$Ra, 2 mBq of $^{40}$K and 0.2 mBq of $^{60}$Co.
  We also examined and resolved the intrinsic leakage of Xe
  gas into PMTs that was observed in several older models.  \\
  We thus succeeded in developing a PMT that has low background,
  large angular acceptance with high collection
  efficiency, good timing resolution, and long-term stable operation.
  These features are highly desirable for experiments searching for rare events
  beyond the standard model, such as
  dark matter particle interactions and neutrinoless double beta decay events.

 }

\arxivnumber{1234.56789} 

\maketitle

\section{Introduction}
\label{sec:intro}
In modern astroparticle physics, one of the foremost scientific challenges is
to directly detect the dark matter interactions with detector materials.
Weakly Interacting Massive Particles (WIMPs) are one of the leading dark matter candidates and
they would produce a signal via elastic scattering on a target nucleus.
Liquid xenon (LXe) is widely used for dark matter searches, owing
to its well-suited material properties for low-background (BG)
searches, such as high light yield, high density, good stopping power.
Many experiments have reported results of dark matter searches using LXe \cite{Abe:2013tc,Akerib:2012ys,Cao:2014jsa,Aprile:2017aty}.

The expected signal rate for LXe dark matter experiments currently is known to be at most
one event/ton/year.
Since the amount of BG directly impacts the sensitivity of
any dark matter search,
the BG event rate from the detector components must be as low as possible. 
With proper shielding against external radiation,
radioactive impurities (RIs) in the 
detector materials have generally becomes the most important source of BG in 
direct-detection dark matter experiments.
Such experiments typically require a large number of photomultiplier 
tubes (PMTs), which often are the largest contributor to the RI budget
in a detector.
This was also the case for the XMASS-I experiment, which carried out many rare-event searches including searches for WIMP dark matter \cite{Suzuki:2018xek,Kobayashi:2018jky,Abe:2018owy,XMASS:2018bid,Abe:2018mxq,Abe:2012az}.
The R10789 PMTs were the one of main BG sources in the experiment,
even though they had been specifically developed for
the XMASS-I experiment \cite{Abe:2018biz}.

New types of sensors, such as silicon photomultipliers, are being rapidly developed, however
the PMT still has the main role for upcoming experiments such as XENONnT \cite{Aprile:2017aty} and LZ \cite{Mount:2017qzi}.
To achieve higher sensitivity in future larger LXe detectors,
it is crucial to develop PMTs with still lower BG; lower RIs and better performance. 
Even after all the effort to date, including the R10789 development,
this is still true.
Such improved PMTs will not only be needed to improve dark matter searches,
but also be useful in other rare event searches
beyond the standard model
such as search for neutrinoless double beta decay.
Based on the R10789 PMTs used in the XMASS-I experiment,
we developed the new R13111 low-BG PMT with good performance and ultra-low RI contamination.
To reduce the RIs, we screened in particular those parts that contributed most to the total RI of the R10789 PMT. 

We adopted a convex-shaped photocathode for the detection of photons incident
on the photocathode from side direction of the PMT. 
When a PMT array is used in a single-phase liquid noble gas detector, 
this allows effective identification of surface BG events
by detecting scintillation photons from the surface on which the PMTs are arranged. 
This property is of particular interest for us,
since the R10789 with its flat photocathode was not suited for the identification
of BG events at these surface.
The surface BG was a very significant BG in XMASS-I \cite{XMASS:2018bid}.

A problem with Xe leakage and the corresponding Xe afterpulses were reported
in \cite{Barrow:PMT, Li:PMT} 
for the R11410 Hamamatsu low-RI PMTs employed in recent LXe detectors
\cite{ Cao:2014jsa, Aprile:2015pmt}.
We also carried out a detailed study of this leakage problem.
In addition to the fact that the afterpulses are adding BG,
a large Xe leak ultimately makes the PMT inoperable. 
Since the R11410 PMT has many structures in common with other Hamamatsu low-BG PMT, Xe leakage is
a problem that needs to be solved
for the long-term stability of all these low-BG PMTs.

In the following section, 
we discuss the structure and performance of the R13111.
In section \ref{sec:RIreduction}, we explain the RI reduction in the PMT,
including our methods for RI measurements, key items for RI reduction, 
and report the results. 
We also discuss the ultra-low BG, high-purity germanium (HPGe) detectors
used to measure the RI of the assembled PMT with high sensitivity
in section \ref{sec:RIreduction}.
We describe the Xe leak study in section  \ref{sec:Afterpulse}.
Our conclusions are given in section  \ref{sec:Conclusion}.

\section{The R13111 PMT}
\subsection{The structure of the PMT}
\label{sec:structure}
\begin{figure}\centering
  \includegraphics[width=9cm]{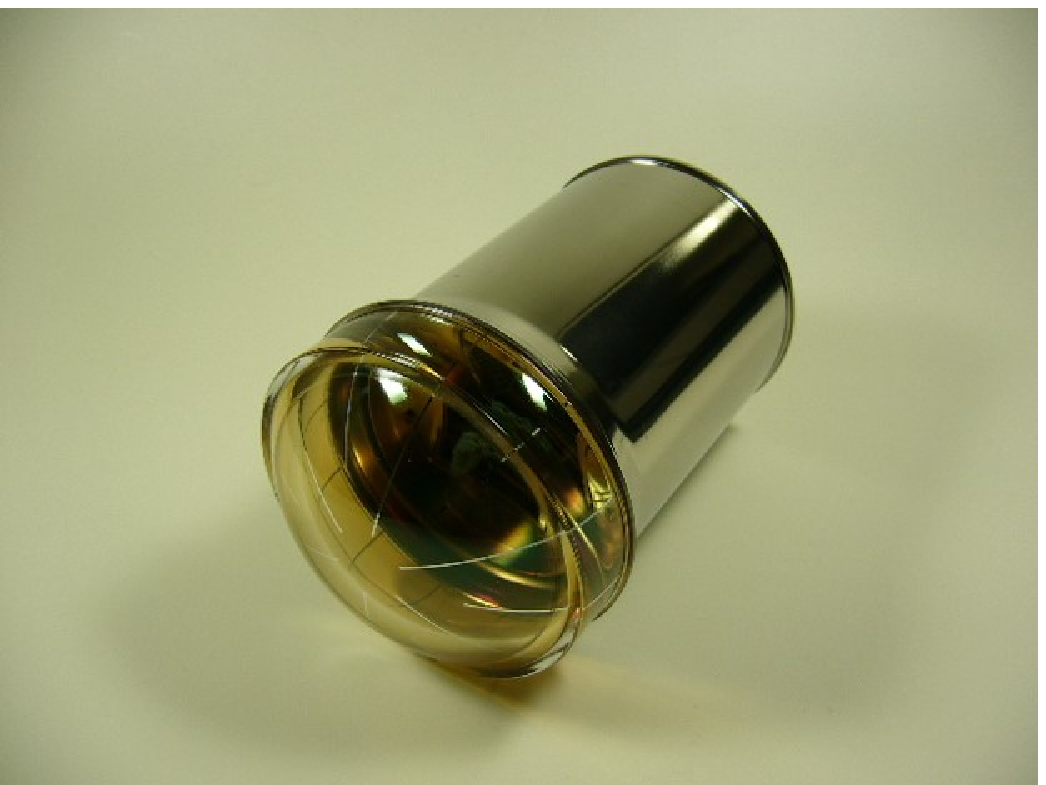}
  \caption{R13111}
  \label{fig:R13111}
\end{figure}

\begin{figure}\centering
  \includegraphics[width=14cm]{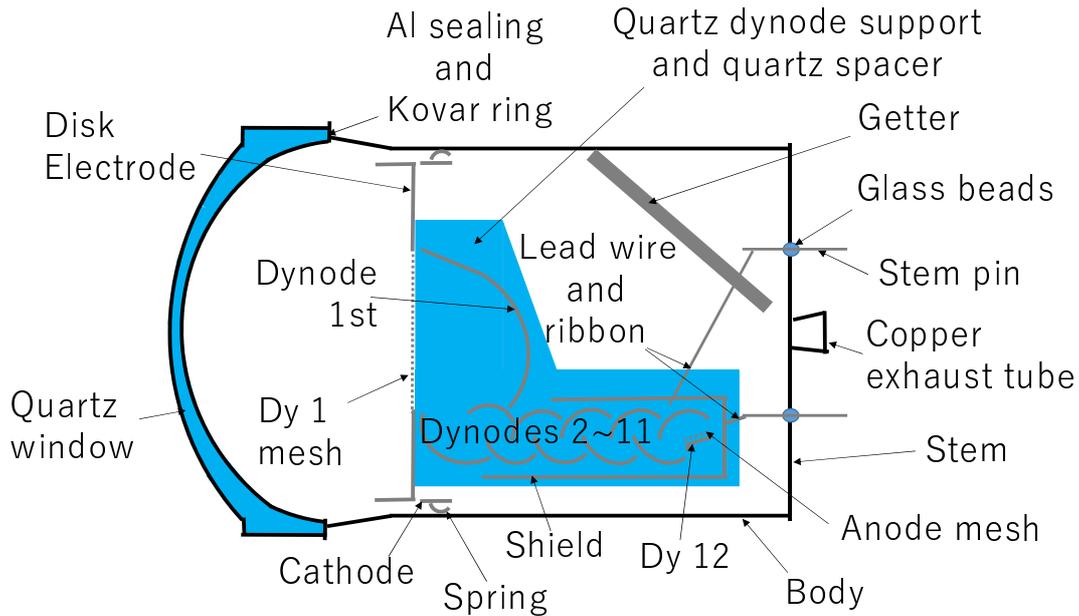}
  \caption{Schematic view of R13111. The disk stopper, the disk feedthrough, the number plate, the screws, the nut and the washer are attached to the disk electrode.}
  \label{fig:pmtfig}
\end{figure}

A picture of an R13111 PMT is shown in Figure \ref{fig:R13111}. 
Figure \ref{fig:pmtfig} shows a schematic view of this R13111 PMT. 
We designed it based on the R10789, which has a smaller diameter of two inches.
Considering this PMT's potential for use in future large-scale detectors, we adopted
a three inch diameter for the R13111.
As with other PMTs used in  LXe detectors, the R13111 PMT uses a quartz window,
which is low in RIs and has high transmittivity for the VUV light from LXe scintillation.
Table \ref{tab:rpmts} lists all the PMT parts with their respective weight
contribution to one R13111 PMT;
their locations in the PMT structure can be seen in Figure \ref{fig:pmtfig}.
A quartz spacer is used to support the dynode structure.
The disk feedthrough is a small feedthrough including the glass part used to feed current
through the disk electrode. 
A screw, nut, and washer attach this disk feedthrough to the disk electrode.
A lead wire and ribbons provide connections between the electrodes and the stem pins
at the back of the PMT.
A spring is used to connect the cathode to the PMT body.

We adopted a convex-concave shape for the R13111 window.
The photocathode is deposited on the window's inside concave surface.
Because of this photocathode geometry,
the R13111 PMT has a large angular acceptance.
This is a main structural difference to the R10789, which has a flat photocathode.
In addition to this photocathode shape,
high collection efficiency for the whole photocathode region
was achieved by designing the geometry of the body parts
and the dynodes using photoelectron-tracking simulations conducted by the manufacturer. 
The design collection efficiency is more than 90\%
for 97\% of the total photocathode area.

The convex geometry of the photocathode also improves the timing resolution,
since this geometry makes the photoelectron track length more uniform over the photocathode area.
The design of the PMT body and its dynodes also are optimized for timing resolution.
This improvement enables better particle identification through the decay time of
the scintillation signal, which directly translates into better BG suppression.

\begin{table}[htb]
  \caption{List of R13111 parts, including their respective weight per PMT.
    The ratio of the weight used for the HPGe measurements to the weight per PMT is also shown.}
  \scalebox{0.9}{
  \begin{tabular}{|c|r|r|r|r|}
    \hline
    Samples & Materials & Weight for & Weight per   & Ratio\\
    & &   HPGe (g) & PMT (g)  &\\
    \hline
       Body & Co free metal & 4450 & 84.9  & 52.4\\
       Stem & Co free metal & 3557 & 62   & 57.4\\
       Stem pin &   Kovar alloy & 92 & 1.6   & 58\\
       Glass beads     &   Made from high-purity glass   & 186 & 2.3  & 80.8 \\
       Quartz window &  Synthetic quartz   & 2610 & 82.1   & 31.8\\
       Quartz dynode support  &  Synthetic quartz   & 360 & 11.8   & 30.5\\
       Quartz spacer        &  Synthetic quartz   & 37.5 & 0.75   & 50\\
       Al Seal    &   Aluminum   & 109.6 & 0.9   & 121.8\\
       Kovar ring    &   Kovar  alloy & 196 & 3.3   & 59.4\\
       Antimony for evaporation     &   Antimony   & 1.6 & 0.022  & 72.7 \\
       Al for evaporation     &   Aluminum   & 19.9 & 0.064   & 311\\
       Hermetic       &   glass, Kovar    & 103 & 0.19   & 542\\
       Lead wire, Ribbons       &   Nickel    & 189 & 1.88   & 101\\
       Screw, nut, washer, getter stopper        &   Non-Magnetic (NM) stainless steel    & 115 & 1.2  & 95.8 \\
       Cathode        &   NM stainless steel (0.25 t)   & 133.5 & 2.4  & 55.6 \\
       Dynode 1st stage        &   NM stainless steel  (0.3 t)    & 212 & 5.1 & 41.6  \\
       Dynode (2nd to 12th stage)        &   NM stainless steel  (0.2 t)    & 375 & 7.5 & 50  \\
       Shield      &   NM stainless steel  (0.3 t)    & 331 & 6.6  & 50.1 \\
       Spring, number plate       &   NM stainless steel      & 32.4 & 0.34 & 95.3  \\
       Anode mesh      &   NM stainless steel (0.25 t)     & 15.6& 0.1   & 156 \\
       Disk stopper      &   NM stainless steel (0.18 t)     & 41.4 & 0.87  & 47.6 \\
       Dynode 1st stage mesh & SUS304 (0.05 t) & 8.1 & 0.16  & 50.6\\
       Disk      &   NM stainless steel (0.5 t)     & 654 & 12.6  & 51.9 \\
       Getter      &  Zr-Al alloy   & 7.0 & 0.07   & 100\\
       Exhaust tube &   Copper   & 86 & 1.4   & 60\\

          \hline
  \end{tabular}
  }
  \label{tab:rpmts}
\end{table}

\subsection{Performance of the PMT}
\label{sec:Performance}
The acceptance, collection efficiency and timing resolution of the R13111 PMT,
were measured with the methods described in \cite{Xu:2016pha}.

At room temperature, a total of 11 PMTs were measured.
The collection efficiency times the quantum efficiency was examined and evaluated relative
to the center of the photocathode.
Observed relative efficiencies were above 70\% at all test points \cite{Xu:2016pha}.
Furthermore, we evaluated the transit time spread (TTS).
The TTS for the whole photocathode area was measured using diffused 405 nm laser light with 63 psec pulse width (FWHM) 
on three of the R13111 PMT. The measured TTS (FWHM) from the three PMTs was 2.11 $^{+0.31}_{-0.18}$ nsec.
Considering that TTS becomes shorter at longer incident wavelength, this number
is consistent with the design value, 2.9 nsec for 175nm,
the Xe scintillation wavelength.
This constitutes a significant improvement over the R10789's TTS of 6.87 $\pm$ 0.31 nsec,
which was obtained using the same setup.

\section{Reduction of RI}
\label{sec:RIreduction}
In this section, we report the achieved reduction of RI.
We first describe the methods used for the RI measurements.
Next we discuss the key materials targeted for RI reduction.
We also report the RIs for these individual materials as well as for the whole assembled PMT.
Finally we summarize the measured RIs and compare them to those reported for other PMTs.

\subsection{Methods used in our RI measurements}
\label{sec:Methods}
To measure the radioactivity of materials we used four different
low BG HPGe detectors  manufactured  by CANBERRA and installed underground at Kamioka.
We also employed two different mass-spectrometry methods; inductively coupled plasma mass
spectrometry (ICP-MS) and glow-discharge mass spectrometry (GD-MS).
Three of the  HPGe detectors are coaxial p-type devices. The other is a coaxial n-type.
The crystal masses and efficiencies of these HPGe detectors are about 2.5 kg/120\% (p-type, GC series), 1.9 kg/100\% (p-type GC series), 1.7 kg/80\% (p-type, T2FA series), and  1.9 kg/100\% (n-type GR series)
\footnote{
  The efficiency is given relative to the efficiency of
  a three-inch diameter 
  three-inch thick NaI(Tl) for 1.332 MeV gamma-ray from  $^{60}$Co source
  which is positioned 25 cm away from the detector.}
.
Details of the HPGe detectors (except for the T2FA series HPGe detector, which we describe below),
the ICP-MS, and the GD-MS 
are provided in \cite{Abe:2018biz}.

\begin{figure}\centering
  \includegraphics[width=12cm]{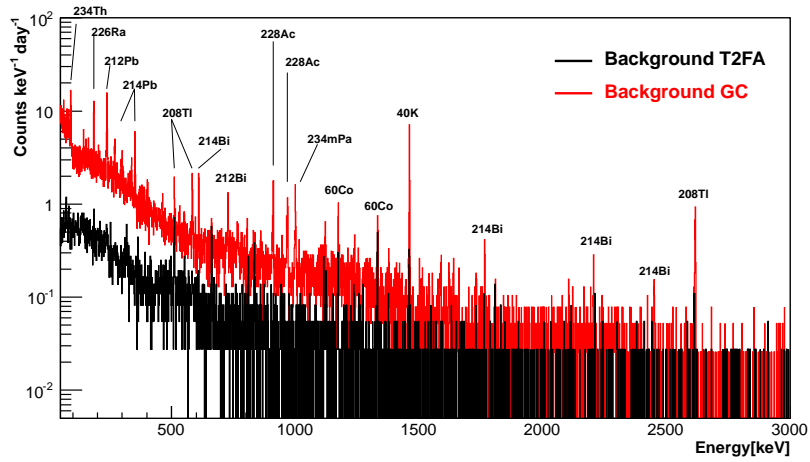}
  \caption{BG spectrum of the T2FA p-type HPGe detector (36 live days, black),
    together with the BG spectrum of a GC p-type HPGe detector (38 live days, red).
    The bin width is 1 keV. }
  \label{fig:HPGeBG}
\end{figure}

To measure low radioactivity with high sensitivity, especially for fully assembled PMTs,
we installed a T2FA series HPGe detector in 2016.
This detector was manufactured by CANBERRA France.
At Kamioka it is located at the center of a  23 $\times$ 23 $\times$ 48 cm sample chamber. To shield it from environmental gamma-rays, this chamber is surrounded by 1 cm of 6N grade copper, 5 cm of OFHC copper, a 2.5 cm lead layer with a $^{210}$Pb activity of only 5 Bq/kg, and a 20 cm lead layer with a $^{210}$Pb activity of 140 Bq/kg.
Like for the other three Kamioka HPGe detectors, this sample chamber is continually flushed with Rn-free air in which the $^{222}$Rn concentration was less than a few tens of mBq/m$^{3}$
in order to suppress radon diffusion into the chamber. 
The sample space in this detector is large enough to measure 13 PMTs at once, while our GC type HPGe detectors can only hold 2 PMTs.
The signal from this HPGe detector was fed into an ORTEC 572A main amplifier.
The amplified signal was recorded by a CAEN DT5724 waveform digitizer with
a 100 MHz sampling rate, which on another channel also recorded the same
signal without amplification.
The maximum pulse height is used to estimate the deposited energy.  Noise can be rejected based on pulse shape.
Detector calibrations were frequently performed with radioactive sources of $^{60}$Co, $^{133}$Ba,
and $^{137}$Cs.  
In order to evaluate detection efficiencies, we performed Geant4 \cite{Agostinelli:2002hh}
based Monte Carlo simulations.  
Similar to our procedure for the other HPGe detectors reported in \cite{Abe:2018biz},
we estimated systematic uncertainties for detection efficiencies 
through comparison between calibration data and simulation 
to be +30\% and -10\%. 
Since 
the systematic error is common to all measurements,
the results
in the following sections are given only with
their respective
statistical errors.
Figure \ref{fig:HPGeBG} shows the BG spectrum of this new detector
together with one from one of the other coaxial p-type HPGe detectors in Kamioka.
Thanks to its ultra-low BG and its large sample chamber, the sensitivity to $^{226}$Ra and $^{228}$Ra are improved from $\mathcal{O}(10^{3})$ $\mu$Bq/PMT for the GC type Ge detectors
to $\mathcal{O}(10^{2})$ $\mu$Bq/PMT for the new T2FA type Ge detector.

GD-MS and ICP-MS measurements were carried out by
Eurofins EAG Materials Science Tokyo Corporation\cite{MS:nanos} and
Toray Research Center, Inc.\cite{MS:toray}.
The systematic uncertainties caused by processes such as
weighing, pretreatment and the measurement itself 
are estimated as $\pm$ 30\% for GD-MS results
and $\pm$10\% for ICP-MS results from standard samples and repeated measurements.

\subsection{Key items for RI reduction}
We concentrated our efforts at RI reduction on those items
that contributed most to the radioactivity of the R10789 PMT\cite{Abe:2018biz}: 
the glass beads used to seal the electrical feedthroughs in the base of the PMT,
the potassium compound used in the photocathode production, and the aluminum seal
for the entrance window.
As a result of this effort 
we achieved a significantly lower RI contamination for the new R13111 PMT.

\subsubsection{The glass beads}
Glass beads are used as electrical insulation for the feedthroughs (Figure \ref{fig:pmtfig}),
and they contributed about 70\% of the total $^{226}$Ra activity in the R10789 PMT. 
Since the total amount of glass needed is not large 
(2 g/PMT as shown in Table \ref{tab:rpmts})
we decided that the glass beads should be made from high-purity reagents. 
The new glass was synthesized with the cooperation of OKAMOTO GLASS Co, Ltd. 
We determined the required reagents and their amounts by analyzing the glass beads used for the R10789,
obtaining the results listed in Table \ref{tab:glassbeads2015}.
High purity reagents were obtained from Kojundo Chemical Laboratory Ltd.
After checking the RIs of all reagents, the synthesis and milling processes were carried out by OKAMOTO GLASS Co, Ltd.
Then, the glass powder was molded and melted into glass beads by the manufacturer.
 
Table \ref{tab:glassbeads2015} shows the amount and measured RI for each reagent,
the synthesized glass,
the glass after the milling process, and the glass beads, together with the measurements
for the glass beads used in the R10789.
Compared to the R10789 glass beads, the RIs of the newly synthesized glass beads are reduced
to less than 2\%, less than 22\%, less than 7\% and 12\% 
for $^{226}$Ra, $^{238}$U, $^{232}$Th and $^{40}$K, respectively.
We measured the RIs of the glass just after synthesis as well as of the milled powder
in order to check for any unexpected
contamination introduced in individual processing steps.
We observed no significant contamination or increase above expectation from the reagents
used to synthesize the glass.
The synthesized glass shows smaller amounts of $^{226}$Ra than the sum of the reagents,
though the measurements could be considered compatible given their uncertainties.
Since the yield of the synthesis was not 100\%, we cannot deny the possibility of such a reduction occurring during the synthesis
process.
The RIs of the glass after milling are higher than those of the glass after synthesis.
We speculate that there was contamination from the air or from the milling tools during the milling process. 
The milling process uses ceramic tools (ceramic balls and a ceramic pot).
An ordinary ceramic tool we checked included 
(3.76 $\pm$ 0.02) $\cdot$ 10$^{3}$  $\mu$Bq/g of $^{226}$Ra,
(4.74 $\pm$ 0.02)  $\cdot$ 10$^{3}$  $\mu$Bq/g of  $^{228}$Ra and (12.0 $\pm$ 0.1)$\cdot$ 10$^{3}$  $\mu$Bq/g of $^{40}$K.
To lower their impact, we also prepared high-purity ceramic tools
(Nikkato, alumina ball SSA-999W and alumina pot mill SSA-999W A-5).
The RIs of the new tools were 
(4.2 $\pm$ 0.1) $\cdot$ 10$^{2}$  $\mu$Bq/g of  $^{226}$Ra,
(35 $\pm$ 3) $\mu$Bq/g of $^{228}$Ra and (68 $\pm$ 22) $\mu$Bq/g of $^{40}$K.
These numbers are one to two orders of magnitude smaller than the ordinary tools, however they
are still quite large compared to the glass beads themselves (see Table \ref{tab:glassbeads2015}).
The possibility of contamination from the tools cannot be rejected.

\begin{table}[htb]
  \caption{Results of HPGe measurements for the glass-bead-related materials. The unit is $\mu$Bq/g.
  Errors quoted are purely statistical. See the text for the systematic error.}
  \scalebox{0.7}{
  \begin{tabular}{|c|r|r|r|r|r|r|r|}
    \hline
    Samples & component & $^{226}$Ra  & $^{238}$U & $^{210}$Pb & $^{228}$Ra & $^{40}$K & $^{60}$Co \\
    & ratio [\%]&& ($^{234}$Th) & & & & \\ \hline
       SiO$_{2}$  5N grade&64   & < 11 & < 1.5 $\cdot$ 10$^{2}$   &  < 1.1 $\cdot$ 10$^{3}$  
       		&  < 13 &   < 1.5 $\cdot$ 10$^{2}$ &  < 4.4   \\ \hline

B$_{2}$O$_{3}$   4N5 Grade  & 3  & < 45 & < 1.8 $\cdot$ 10$^{2}$   &  < 4.6 $\cdot$ 10$^{2}$  
       		&  < 44  &   < 4.5 $\cdot$ 10$^{2}$ &  < 12   \\ \hline
		
Na$_{2}$B$_{4}$O$_{7}$   4N5 Grade  & 24 & (1.8 $\pm$ 0.6) $\cdot$ 10$^{2}$ & < 3.9 $\cdot$ 10$^{2}$   &  < 1.3 $\cdot$ 10$^{3}$  
       		&  < 74  &   < 1.1 $\cdot$ 10$^{3}$ &  < 25  \\ \hline

       Al(OH)$_{3}$  4N grade& 9 & < 25 & < 2.3 $\cdot$ 10$^{2}$   &  < 1.5 $\cdot$ 10$^{3}$  
       		&  < 42  &   < 2.0 $\cdot$ 10$^{2}$ &  < 7.6   \\ \hline
   \hline
   Sum &   & 36 $\pm$ 15   &< 1.3 $\cdot$ 10$^{2}$ & < 6.5 $\cdot$ 10$^{2}$ 
   		 & < 19  & < 2.6 $\cdot$ 10$^{2}$ & < 6.6  \\  \hline
   \hline    \hline
   
   Glass  after synthesis & & 12 $\pm$ 4  & 57 $\pm$ 26  & < 61 
   				& < 9.4  &   (1.5 $\pm$ 0.5 )  $\cdot$ 10$^{2}$  &  < 2.5  \\ \hline
	 
  Glass  after milling & & 24 $\pm$ 5   & < 97  & < 99 
   				& 17 $\pm$ 4  &   (1.3 $\pm$ 0.6 )  $\cdot$ 10$^{2}$  &  < 4.0  \\ \hline
				
  Glass  beads & & <  15  & < 72  & < 1.9 $\cdot$ 10$^{2}$ 
   				& < 9.7  &   (1.7 $\pm$ 0.7 )  $\cdot$ 10$^{2}$  &  < 2.0  \\ \hline
  \hline \hline
				
 R10789 glass beads  & & (8.1 $\pm$ 0.1)  $\cdot$ 10$^{2}$ & (3.3 $\pm$ 0.6)  $\cdot$ 10$^{2}$   & < 4.1 $\cdot$ 10$^{2}$ 
   				& (1.4 $\pm$ 0.1) $\cdot$ 10$^{2}$ &   (1.4 $\pm$ 0.1 )  $\cdot$ 10$^{3}$  &  < 2.3  \\ \hline	
  \end{tabular}
}  
  \label{tab:glassbeads2015}
\end{table}

\subsubsection{The potassium compound used for photocathode production}
Most of the $^{40}$K in an assembled R10789 is presumed to come from the photocathode material,
which is deposited on the back of the entrance window 
during the assembly process
\cite{Abe:2018biz}.
Chemicals used to make the photocathode material contain potassium,
which contains the radioactive isotope $^{40}$K with a natural abundance of 117 ppm.
Since potassium is an essential ingredient for the photocathode,
$^{39}$K-enriched potassium chromate from ATOX Co., Ltd. was used to produce the photocathode for the
new R13111 PMT.
Table \ref{tab:PhotocathodeMaterial} shows the
RIs of two samples of potassium chromate,
one without enrichment and the other enriched in $^{39}$K.
For the one without enrichment, the amount of $^{40}$K was 10 $\pm$ 1 Bq/g. This corresponds to 99.8 $\pm$ 0.7 ppm,
which is consistent with the natural abundance within the systematic error
of the HPGe measurements.
In contrast $^{39}$K-enriched potassium chromate has 84 $\pm$ 4 mBq/g, or 0.81 $\pm$ 0.04 ppm
of $^{40}$K, a reduction of more than two orders of magnitude.

\begin{table}[htb]
  \caption{Results of HPGe measurements for 20 g of the photocathode material. The units are $\mu$Bq/g.
  Errors in the table include only statistical error. See the text for the systematic error.}
  \scalebox{1.0}{
  \begin{tabular}{|c|r|r|r|r|r|r|}
    \hline
    Samples & $^{226}$Ra  & $^{238}$U & $^{210}$Pb & $^{228}$Ra & $^{40}$K & $^{60}$Co \\
    && ($^{234}$Th) & & & & \\ \hline
  without enrichment & < 82  & < 3.0 $\cdot$ 10$^{4}$ & < 8.8 $\cdot$ 10$^{4}$ & < 3.4 $\cdot$ 10$^{2}$  & (1.0 $\pm$ 0.1) $\cdot$ 10$^{7}$ & < 9.9 $\cdot$ 10$^{2}$ \\ \hline
  with enrichment & < 2.2 $\cdot$ 10$^{2}$  & < 1.3 $\cdot$ 10$^{3}$ & < 8.9 $\cdot$ 10$^{3}$ & < 1.9 $\cdot$ 10$^{2}$  & (8.4 $\pm$ 0.4) $\cdot$ 10$^{4}$ & < 52 \\ \hline
  \end{tabular}
  }
  \label{tab:PhotocathodeMaterial}
\end{table}

\subsubsection{The aluminum seal}
Aluminum is used for the seal between the quartz window and the metal body of both the R10789 and the R13111.
For the R10789,  98.5\%
purity aluminum was used for the seal.
During the XMASS 800kg commissioning phase, we found that a large amount of the surface BG was  
caused by radioisotopes included in this aluminum seal.
The relevant isotopes came from the top part of the $^{238}$U decay chain
and $^{210}$Pb \cite{Abe:2013tc}.
For the R13111, we decided to use higher-purity, 5N aluminum (Nippon Light Metal Company, LTD.)
for the seal. This reduced
these isotopes substantially as shown in Table \ref{tab:Alseal}.
The amount of $^{238}$U measured by mass spectrometry shows
three orders of magnitude reduction.
For $^{210}$Pb, more than about two orders of magnitude reduction was achieved.
Other RIs are also reduced considerably by the increased purity of the new material.

\begin{landscape}
\begin{table}[htb]
  \caption{Results of HPGe and mass spectrometry measurements for the aluminum seal materials. The units are $\mu$Bq/g. Errors for HPGe measurements in the table include only statistical error. See the text for the systematic error.}
  \scalebox{0.85}{
  \begin{tabular}{|c|r|r|r|r|r|r||r|r|r|}
    \hline
    Samples & $^{226}$Ra  & $^{238}$U & $^{210}$Pb & $^{228}$Ra& $^{40}$K & $^{60}$Co & $^{238}$U & $^{232}$Th  & Method for  \\
                   &                      &   &                 &               &      &      &&                         &    $^{238}$U and $^{232}$Th  \\ \hline
    used in 2015 Production (5N purity)       & <  10  &  < 51  & < 1.0 $\cdot$ 10$^{2}$   &  13 $\pm$ 5 & < 1.8 $\cdot$ 10$^{2}$
    		&  < 5.8  &  2.5  & 4.1 $\cdot$ 10$^{-1}$ & GD-MS \\ \hline
    used in 2016 Production  (5N purity)     & <  2.8  &  < 37  & < 66  &  < 2.5  & < 34
    		&  < 9.6 $\cdot$ 10$^{-1}$ &  2.5  & 4.1 $\cdot$ 10$^{-1}$ & GD-MS \\ \hline
    used in R10789 PMT  (98.5\% purity)     & <  36 &  (3.3 $\pm$ 0.4)  $\cdot$ 10$^{3}$  & (7.7 $\pm$ 6.0) $\cdot$ 10$^{3}$   &  (1.0 $\pm$ 0.3 ) $\cdot$ 10$^{2}$  &< 7.1 $\cdot$ 10$^{2}$
    		&  < 18  &  3.9 $\cdot$ 10$^{3}$ & 74  & ICP-MS \\ \hline

      \end{tabular}
  }
\label{tab:Alseal}
\end{table}
\end{landscape}

\subsection{Resulting improvements}
\subsubsection{RIs of the PMT components}
\label{sec:RIcomp}

We report the RIs in each component of the newly developed R13111 in this section.
Table \ref{tab:rpmts} shows a list of components measured by the HPGe detectors, 
together with the weight of each part (measured during the year 2015).
We prepared large
samples
which were 30-500 times larger than the amount of material used for a single PMT
in order to have the required sensitivity.

The amounts of $^{226}$Ra, $^{238}$U and $^{210}$Pb from the $^{238}$U decay chain,
$^{228}$Ra from the $^{232}$Th chain, and $^{40}$K and $^{60}$Co were determined
using the HPGe detectors. 
Mass-spectrometry was used to measure the amounts of  $^{238}$U, $^{232}$Th,
and $^{40}$K assuming the natural abundance of
isotopes for each element. 
These mass-spectrometry measurements were performed to increase
the sensitivity for materials with large volumes per PMT.
Mass spectrometers have better sensitivity than HPGe,
which becomes relevant when large total amounts of material increase the impact of
even tiny contaminations.

The results of the measured RIs for PMTs assembled in years 2015 and 2016 are shown
separately in Table \ref{tab:rpmt2015}
and Table \ref{tab:rpmt2016}, respectively.
In case the center value does not exceed twice the statistical error, or is negative,
we regarded the result as consistent with zero and set an upper limit calculated as
$\max\{0, {\mbox{the central value}}\}+1.28\times \mbox{statistical error}$.  
Though the PMTs produced in both years used the same materials from the same companies,
the observed numbers are different for many parts.
This is because the production lots of those parts are not the same. 
For $^{226}$Ra, $^{210}$Pb, $^{228}$Ra, and $^{60}$Co, 
the total activities of the PMT components are also shown in Tables
\ref{tab:rpmt2015} and \ref{tab:rpmt2016}.
In calculating of the total radioactivity, we summed up the central values of each
component, with the sign of each value included in the sum.
To calculate the total activity of $^{238}$U and $^{40}$K,
we used the mass spectrometer results instead of HPGe results,
if the mass spectrometer results exist and the sensitivity of the results were
better than that of the HPGe result.  
These are shown in Table \ref{tab:comparison} as Part-by-part sum (2015) and
Part-by-part sum (2016).
Since the errors in the table show only statistical error,
systematic errors would add another +30\% to -10\% to the HPGe, 
$\pm$ 30\% to the GD-MS, and $\pm$ 10\% to the ICP-MS measurements results.

Among the components in Tables \ref{tab:rpmt2015} and \ref{tab:rpmt2016},
some stainless-steel parts 
in particular bring large contributions
of $^{226}$Ra, $^{228}$Ra and $^{60}$Co compared to the total. 
Most prominent are the cathode, the disk, the 1st dynode, the 2nd-11th dynodes,
and the shield.
Such stainless-steel parts
also differ noticeably in their respective
 2015 and 2016 RI contributions. 
To manufacture them, we used commercial products that have large variations in the RIs.
This also applies for the cobalt-free metal used in the PMT body and the stem during 2015,
which have
large values of $^{228}$Ra and $^{60}$Co, while the same material used in 2016 shows smaller amounts.
The glass parts (the glass beads and the disk feedthrough) still have large amounts of
$^{226}$Ra, $^{228}$Ra and especially $^{40}$K, though we largely reduced the RIs in the glass beads.

\begin{landscape}
\begin{table}[htb]
  \caption{Result of the measurement for each part used for R13111 production in 2015. The units are $\mu$Bq/PMT.
  Errors in the table include only statistical error. See the text for the systematic error.}
  \scalebox{0.75}{
  \begin{tabular}{|c|r|r|r|r|r|r||r|r|r|r|}
    \hline
     & \multicolumn{6}{|c||}{results of HPGe   } & \multicolumn{4}{|c|}{result of the mass spectrometry}\\ \hline
    Samples & $^{226}$Ra  & $^{238}$U & $^{210}$Pb & $^{228}$Ra& $^{40}$K & $^{60}$Co & $^{238}$U & $^{232}$Th & $^{40}$K & Method for  \\
                   &                      &   &                 &             &      &      &&                         &   &  $^{238}$U and $^{232}$Th  \\ \hline

Body   & < 1.0  $\cdot$ 10$^{2}$ &  < 3.1 $\cdot$ 10$^{3}$  & < 1.2 $\cdot$ 10$^{5}$   &  (1.2 $\pm$ 0.4) $\cdot$ 10$^{2}$  &< 5.5 $\cdot$ 10$^{2}$
    		&  (1.5 $\pm$ 0.2) $\cdot$ 10$^{2}$ &  2.1 $\cdot$ 10$^{2}$ & < 17 & -- & GD-MS \\ \hline 

Stem   & < 1.0  $\cdot$ 10$^{2}$ &  < 3.0 $\cdot$ 10$^{3}$  & < 8.8 $\cdot$ 10$^{4}$   & < 1.4 $\cdot$ 10$^{2}$  &< 8.4 $\cdot$ 10$^{2}$
    		&  64 $\pm$ 18 &  5.5 $\cdot$ 10$^{2}$ & < 26  & < 2.0 & GD-MS\\ \hline

Stem pin    & < 24  &  < 2.8 $\cdot$ 10$^{2}$  & < 9.2 $\cdot$ 10$^{3}$   & < 27  &< 2.2 $\cdot$ 10$^{2}$
    		&  86 $\pm$ 9  & < 2.0 & < 6.5 $\cdot$ 10$^{-1}$ & < 2.5 $\cdot$ 10$^{-1}$ & GD-MS\\ \hline

Glass beads   & <  34  &  < 1.7 $\cdot$ 10$^{2}$  & < 4.5 $\cdot$ 10$^{2}$   & < 22 & (3.9 $\pm$ 1.6 ) $\cdot$ 10$^{2}$
    		&  < 4.7  & --  & -- & -- & --\\ \hline

Quartz Window    & < 2.0 $\cdot$ 10$^{2}$ & < 1.6 $\cdot$ 10$^{3}$  & < 6.8 $\cdot$ 10$^{3}$ & < 1.7 $\cdot$ 10$^{2}$ & < 2.8 $\cdot$ 10$^{3}$
    		& < 49  & < 20  & < 6.7  & 7.6 $\cdot$ 10$^{-2}$ & ICP-MS\\ \hline
		
Quartz dynode support & < 44  & < 7.9 $\cdot$ 10$^{2}$  & < 3.5 $\cdot$ 10$^{4}$ & < 32 & < 5.3 $\cdot$ 10$^{2}$
    		& < 8.1  & < 3.0  & < 9.8 $\cdot$ 10$^{-1}$ & 7.8  & ICP-MS\\ \hline

Quartz spacer    & < 20  & < 1.4 $\cdot$ 10$^{2}$  & < 7.0 $\cdot$ 10$^{2}$ & < 20 & < 1.4 $\cdot$ 10$^{2}$
    		& < 4.3  & < 1.7 $\cdot$ 10$^{-1}$ & < 5.7 $\cdot$ 10$^{-2}$ & 4.6 $\cdot$ 10$^{-1}$ & ICP-MS\\ \hline

Al seal  & <  9.0  &  < 46  & < 94   &  11 $\pm$ 5   &< 1.7 $\cdot$ 10$^{2}$
    		&  < 5.2  &  2.2  & 3.7 $\cdot$ 10$^{-1}$ & -- & GD-MS \\ \hline

Kovar ring   & < 29 & < 7.2 $\cdot$ 10$^{2}$  & < 2.1 $\cdot$ 10$^{4}$ & < 28 & < 2.3 $\cdot$ 10$^{2}$
    		& 37 $\pm$ 8  &  < 4.6  &  < 1.5  & < 5.7 $\cdot$ 10$^{-1}$ & GD-MS\\ \hline

Sb for evaporation   & 15 $\pm$ 6  &  < 40 & < 94 & < 6.1 &< 1.4 $\cdot$ 10$^{2}$
    		&  < 2.2  &  -- &  --& -- & --  \\ \hline
		
Al for evaporation   & < 4.2  &  < 21 & < 33 & < 5.8  &< 83 
    		&  < 1.1 & 1.6 $\cdot$ 10$^{-1}$  & 2.6 $\cdot$ 10$^{-2}$  & -- & GD-MS \\ \hline

Disk feedthrough   & 11 $\pm$ 2  &  < 45 & < 4.8 $\cdot$ 10$^{2}$   & (5.5 $\pm$ 2.4 )   & 71 $\pm$ 34 
    		&  10 $\pm$ 1  &  -- & -- & -- & --\\ \hline

Lead wire   & 42 $\pm$ 8  &  < 2.9 $\cdot$ 10$^{2}$  & < 6.7 $\cdot$ 10$^{3}$   & < 21 &< 1.1 $\cdot$ 10$^{2}$
    		&  < 4.0  &  -- &  --& -- & --\\ \hline

Screw, nut, washer, getter stopper    & < 12 & < 2.8 $\cdot$ 10$^{2}$  & < 3.6 $\cdot$ 10$^{3}$ & < 25 & < 1.3 $\cdot$ 10$^{2}$
    		& < 3.9  &  1.3 $\cdot$ 10$^{2}$ & 2.8  & 1.1  & GD-MS\\ \hline

Cathode    & (1.3 $\pm$ 0.3 ) $\cdot$ 10$^{2}$ &  < 8.2 $\cdot$ 10$^{2}$  & < 5.8 $\cdot$ 10$^{3}$   &  67 $\pm$ 24 &< 4.6 $\cdot$ 10$^{2}$
    		&  23 $\pm$ 10 & 23  & 11 & 1.7  & GD-MS \\ \hline

Dynode 1st stage   & (1.1 $\pm$ 0.3)  $\cdot$ 10$^{2}$ &  < 1.1 $\cdot$ 10$^{3}$  & < 4.7 $\cdot$ 10$^{3}$   & < 66  &< 3.8 $\cdot$ 10$^{2}$
    		&  < 30 & 1.0 $^\cdot$ 10$^{2}$ & 17 & < 6.6 $\cdot$ 10$^{-1}$ & GD-MS\\ \hline

Dynode (2nd - 12th stage)       & <  55 &  < 8.8 $\cdot$ 10$^{2}$  & < 1.4 $\cdot$ 10$^{4}$   &  < 30 & < 3.7 $\cdot$ 10$^{2}$
    		&  29 $\pm$ 10 &  < 9.3  & < 3.0  & < 1.2   & GD-MS\\ \hline

Shield        & < 52   &  < 9.2 $\cdot$ 10$^{2}$  & < 8.5 $\cdot$ 10$^{3}$   & < 55  &< 5.3 $\cdot$ 10$^{2}$
    		&  < 20 &  1.6 $\cdot$ 10$^{2}$ & 27 & < 1.0  & GD-MS\\ \hline

Spring, Number plate    & <  8.8  &  < 1.2 $\cdot$ 10$^{2}$  & < 1.9 $\cdot$ 10$^{3}$   & < 9.1  &< 1.2 $\cdot$ 10$^{2}$
    		&  < 3.5  & 1.8   & 9.4 $\cdot$ 10$^{-1}$ &  < 4.4 $\cdot$ 10$^{-1}$ & GD-MS\\ \hline
		
Anode mesh   & < 16 &  < 1.4 $\cdot$ 10$^{2}$  & < 4.7 $\cdot$ 10$^{3}$   & < 11  &< 2.3 $\cdot$ 10$^{2}$
    		&  < 3.3 &   -- &  -- & -- & --\\ \hline

Disk Stopper   & <  12 &  < 6.7 $\cdot$ 10$^{2}$  & < 1.5 $\cdot$ 10$^{4}$   &  < 14 &< 1.7 $\cdot$ 10$^{2}$
    		&  < 4.5 &  -- & -- &  -- & -- \\ \hline

Dynode 1st stage mesh     & < 22 &  < 89 & < 2.7 $\cdot$ 10$^{2}$   & < 19 & < 2.4 $\cdot$ 10$^{2}$
    		&  < 5.1  &  -- &--  & -- & --\\ \hline

Disk        & 95 $\pm$ 36  &  < 9.5 $\cdot$ 10$^{2}$  & < 1.5 $\cdot$ 10$^{4}$   & 94 $\pm$ 35 &< 4.9 $\cdot$ 10$^{2}$
    		&  94 $\pm$ 17 & 1.1 $\cdot$ 10$^{2}$  & 53 & 8.1 & GD-MS \\ \hline

Getter    & <  5.3  &  < 81 & < 6.1 $\cdot$ 10$^{2}$   & < 5.2   &< 1.2 $\cdot$ 10$^{2}$
    		&  < 1.9 &  -- & -- &-- & -- \\ \hline

Exhaust tube   & < 33  &  < 1.4 $\cdot$ 10$^{2}$  & < 1.1 $\cdot$ 10$^{3}$   & < 23 &< 4.5 $\cdot$ 10$^{2}$
    		&  < 4.6  & -- &  --&-- & -- \\ \hline
		
Sum   &  (6.2 $\pm$ 1.7)  $\cdot$ 10$^{2}$ & -- & < 1.6 $\cdot$ 10$^{5}$   &  (4.8 $\pm$ 1.5) $\cdot$ 10$^{2}$  & -- &  (4.8 $\pm$ 0.6) $\cdot$ 10$^{2}$ & --  & -- & -- & -- \\ \hline 
		    
      \end{tabular}
  }
\label{tab:rpmt2015}
\end{table}

\begin{table}[htb]
  \caption{Result of the measurement for each part used for R13111 production in 2016. The units are $\mu$Bq/PMT.
  Errors in the table include only statistical error. See the text for the systematic error.}
  \scalebox{0.75}{
  \begin{tabular}{|c|r|r|r|r|r|r||r|r|r|r|}
    \hline
         & \multicolumn{6}{|c||}{results of HPGe   } & \multicolumn{4}{|c|}{result of the mass spectrometry}\\ \hline
    Samples & $^{226}$Ra  & $^{238}$U & $^{210}$Pb & $^{228}$Ra& $^{40}$K & $^{60}$Co & $^{238}$U & $^{232}$Th & $^{40}$K & Method for  \\
                   &                      &   &                 &                &      &      &&                         &   &  $^{238}$U and $^{232}$Th  \\ \hline
 
Body   & < 55  &  < 2.3 $\cdot$ 10$^{3}$  & < 7.3 $\cdot$ 10$^{4}$   &  < 1.2 $\cdot$ 10$^{2}$  &< 5.1 $\cdot$ 10$^{2}$
    		&  < 36 &  < 73 & < 17 & 55 & GD-MS\\ \hline

Stem   & < 1.1  $\cdot$ 10$^{2}$ &  < 4.1 $\cdot$ 10$^{3}$  & < 7.4 $\cdot$ 10$^{4}$   & < 1.0 $\cdot$ 10$^{2}$  &< 4.8 $\cdot$ 10$^{2}$
    		&  < 33 & 5.5 $\cdot$ 10$^{2}$ & < 26 & < 2.0  & GD-MS\\ \hline

Stem pin    & < 7.1  &  < 2.7 $\cdot$ 10$^{2}$  & < 1.7 $\cdot$ 10$^{3}$   & < 9.2  &< 86 
    		&  84 $\pm$ 5 &  < 1.8  & < 6.0 $\cdot$ 10$^{-1}$ & < 2.3 $\cdot$ 10$^{-1}$ & GD-MS \\ \hline

Glass beads   & 16 $\pm$ 3 &  < 1.5 $\cdot$ 10$^{2}$  & < 5.9 $\cdot$ 10$^{2}$   & 9.8 $\pm$ 3.4 & (6.7 $\pm$ 0.6 ) $\cdot$ 10$^{2}$
    		&  < 1.2  & < 1.2 $\cdot$ 10$^{2}$  & < 41 & < 6.8 $\cdot$ 10$^{2}$ & ICP-MS\\ \hline

Quartz window    & < 87 & < 1.1 $\cdot$ 10$^{3}$  & < 2.5 $\cdot$ 10$^{4}$ & < 54 & < 2.9 $\cdot$ 10$^{2}$
    		& < 28 & < 20 & < 6.7 & 7.6 $\cdot$ 10$^{-2}$ & ICP-MS\\ \hline
				
Quartz dynode support & < 31 & < 2.5 $\cdot$ 10$^{2}$  & < 2.8 $\cdot$ 10$^{3}$ & < 16 & < 1.2 $\cdot$ 10$^{2}$
    		& < 8.4 & < 3.0  & < 9.9 $\cdot$ 10$^{-1}$ & 7.9  & ICP-MS\\ \hline

Quartz spacer    & < 20 & < 1.4 $\cdot$ 10$^{2}$  & < 7.0 $\cdot$ 10$^{2}$ & < 20 & < 1.4 $\cdot$ 10$^{2}$
    		& < 4.3  & < 1.7 $\cdot$ 10$^{-1}$ & < 5.7 $\cdot$ 10$^{-2}$ & 4.6 $\cdot$ 10$^{-1}$ & ICP-MS\\ \hline

Al seal         & <  2.5  &  < 33 & < 60 &  < 2.2  &< 31
    		&  < 8.7 $\cdot$ 10$^{-1}$ &  2.2  & 3.7 $\cdot$ 10$^{-1}$ & -- & GD-MS \\ \hline

Kovar ring   & 21 $\pm$ 6 & < 1.3 $\cdot$ 10$^{2}$  & < 2.6 $\cdot$ 10$^{4}$ & < 9.4  & < 97
    		& 19 $\pm$ 3 &  < 2.2 $\cdot$ 10$^{2}$ &  < 72 & 5.5 & GD-MS\\ \hline

Sb for evaporation   & < 10  &  < 60 & < 2.2 $\cdot$ 10$^{2}$   & < 6.9  &< 1.0 $\cdot$ 10$^{2}$
    		&  < 1.2  & --  & -- &-- & -- \\ \hline
		
Al for evaporation   & < 4.2  &  < 21 & < 33 & < 5.8  &< 83
    		&  < 1.1 & 1.6 $\cdot$ 10$^{-1}$  & 2.6 $\cdot$ 10$^{-2}$ & -- & GD-MS\\ \hline

Disk feedthrough   & 11 $\pm$ 2 &  < 45 & < 4.8 $\cdot$ 10$^{2}$   & 5.5 $\pm$ 2.4  & 71 $\pm$ 34 
    		&  10 $\pm$ 1 &  -- &  --&-- & --\\ \hline

Lead wire, ribbons, getter stopper   & < 25 &  < 2.5 $\cdot$ 10$^{2}$  & < 1.4 $\cdot$ 10$^{3}$   & < 14 &< 4.6 $\cdot$ 10$^{2}$
    		&  < 6.5  &  54 & 1.7 & 9.2 $\cdot$ 10$^{-1}$ & GD-MS \\ \hline

Screw, nut, washer   & < 18 & < 74 & < 3.3 $\cdot$ 10$^{2}$ & < 6.1  & < 89
    		& < 4.6  &  40 & 15 & 2.1 & GD-MS\\ \hline

Cathode    & 5.4 $\pm$ 2.5   &  < 2.4 $\cdot$ 10$^{2}$  & < 4.8 $\cdot$ 10$^{3}$   &  < 6.2  &< 80
    		&  (1.9 $\pm$ 0.9)  &  < 3.0  & 4.9  & < 3.7 $\cdot$ 10$^{-1}$ & GD-MS\\ \hline

Dynode 1st stage   & < 33  &  < 2.5 $\cdot$ 10$^{2}$  & < 8.8 $\cdot$ 10$^{2}$   & < 20 &< 3.1 $\cdot$ 10$^{2}$
    		&  < 14 &  < 2.8 $\cdot$ 10$^{2}$ & < 93 & < 14 & GD-MS\\ \hline

Dynode (2- 11 stage)       & (1.0 $\pm$ 0.4) $\cdot$ 10$^{2}$ &  < 5.5 $\cdot$ 10$^{2}$  & < 2.5 $\cdot$ 10$^{3}$   &  < 51 &< 5.6 $\cdot$ 10$^{2}$
    		&  32 $\pm$ 13 &  < 9.0  & < 3.0 & < 1.1 & GD-MS\\ \hline

Dynode (12th stage)      & < 20 &  < 1.0 $\cdot$ 10$^{2}$  & < 2.4 $\cdot$ 10$^{2}$   &  < 13 &< 1.3 $\cdot$ 10$^{2}$
    		&  < 2.7 &  2.1  & 9.0 $\cdot$ 10$^{-1}$ & 8.6 $\cdot$ 10$^{-1}$ & GD-MS\\ \hline

Shield        & ( 1.0 $\pm$ 0.2 )  $\cdot$ 10$^{2}$ &  < 7.6 $\cdot$ 10$^{2}$  & < 7.7 $\cdot$ 10$^{3}$   & < 26 &< 3.0 $\cdot$ 10$^{2}$
    		&  < 12 & < 4.1 $\cdot$ 10$^{2}$  & < 1.4 $\cdot$ 10$^{2}$ & < 21 & GD-MS\\ \hline

Spring, number plate & <  8.8  &  < 1.2 $\cdot$ 10$^{2}$  & < 1.9 $\cdot$ 10$^{3}$   & < 9.1  &< 1.2 $\cdot$ 10$^{2}$
    		&  < 3.5  & 1.8   & 9.4 $\cdot$ 10$^{-1}$ &  4.4 $\cdot$ 10$^{-1}$ & GD-MS\\ \hline

Anode mesh, Disk Stopper, Dynode 1st stage mesh    & 23 $\pm$ 9 &  < 2.1 $\cdot$ 10$^{2}$  & < 5.3 $\cdot$ 10$^{3}$   &  < 20 &< 2.6 $\cdot$ 10$^{2}$   &  < 3.8  & --  & -- &-- & -- \\ \hline
 
Disk        & < 20  &  < 1.0 $\cdot$ 10$^{3}$  & < 4.2 $\cdot$ 10$^{3}$   & < 17 &< 4.1 $\cdot$ 10$^{2}$
    		&  11 $\pm$ 5 & < 16 & < 5.1  & < 2.0  & GD-MS \\ \hline

Getter    & <  5.3  &  < 81 & < 6.1 $\cdot$ 10$^{2}$   & < 5.2   &< 1.2 $\cdot$ 10$^{2}$
    		&  < 1.9  &  -- & -- &-- & -- \\ \hline

Exhaust tube   & < 11  &  < 2.6 $\cdot$ 10$^{2}$  & < 7.2 $\cdot$ 10$^{3}$   & < 14 &< 2.4 $\cdot$ 10$^{2}$
    		&  < 4.6  & --  &  --& -- & -- \\ \hline

Sum   &  (3.9 $\pm$ 1.0)  $\cdot$ 10$^{2}$ & -- & < 1.1 $\cdot$ 10$^{5}$   &  < 2.0 $\cdot$ 10$^{2}$  & -- &  (1.9 $\pm$ 0.4) $\cdot$ 10$^{2}$ &  -- & --  & -- & --\\ \hline 
    
      \end{tabular}
  }
\label{tab:rpmt2016}
\end{table}
\end{landscape}

\subsubsection{RIs of fully assembled PMTs}
\label{sec:AssembledPMTRI}
We also carried out the RI measurement of PMTs after
production to also evaluate possible 
contamination during the assembling process and asses the overall reduction of $^{40}$K
in the photocathode; 
the $^{39}$K-enriched potassium compound is used only in the photocathode 
and the RIs of the finished photocathode cannot be measured before assembling the PMT.
The effect of the  $^{39}$K enrichment was examined by measuring
two variants of the R13111 PMT:
one with natural potassium
and the other with $^{39}$K-enriched potassium.

\begin{figure}\centering
  \includegraphics[width=12cm]{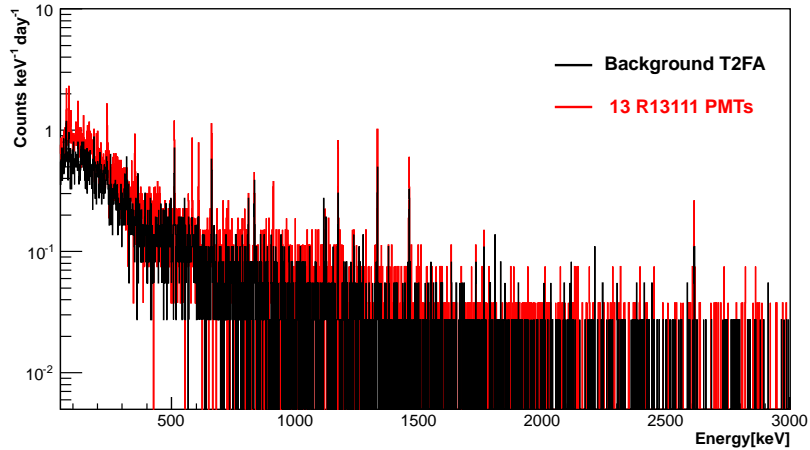}
  \caption{ The energy spectra for 13 R13111 PMTs measured over 27 days (red)
    with the 36 day BG spectra (black).
    For the PMTs produced in 2015 with the $^{39}$K-enriched photocathode material.
    The bin width is 1 keV.
  }
  \label{fig:HPGePMT}
\end{figure}

Figure \ref{fig:HPGePMT} shows the energy spectrum of 13 such PMTs 
of the 2015 variant with the $^{39}$K-enriched photocathode.
Table \ref{tab:comparison} shows the results for the produced new PMT varieties.
In particular, it shows RIs for 2015 with the natural potassium,
2015 with $^{39}$K-enriched potassium, and 2016 with $^{39}$K-enriched potassium. 
The sum of the radioactivity from component values in Table \ref{tab:comparison}
is taken 
from Tables \ref{tab:rpmt2015} and \ref{tab:rpmt2016}.
To calculate the $^{238}$U sum and $^{40}$K sum, we use the mass spectrometry results
if this results exist and the sensitivity of results is better than that of HPGe results.

The results for the PMTs assembled in 2015 with the enriched potassium and
those assembled in 2016 with the enriched potassium
are consistent with each other. Differences are less than two sigma.
However, for each year,
the measurements of RIs for parts often differ from other years
due to the different production lots they were drawn from 
(see section \ref{sec:RIcomp}).
Based on our screening, we controlled the total RI differences which came
from using different lot parts to the level of our measurements.

\begin{table}[htb]
\caption{PMT measurement results. The units are mBq/PMT. Errors in the table include only statistical error. See the text for the systematic error}
\scalebox{0.85}{
\begin{tabular}{|c|r|r|r|r|r|r|}
\hline
mBq/PMT & $^{226}$Ra  & $^{238}$U & $^{210}$Pb & $^{228}$Ra & $^{40}$K & $^{60}$Co  \\ \hline \hline
Assembled PMT (2015) & (4.4 $\pm$ 0.9)  $\cdot$ 10$^{-1}$ & < 2.7 & < 56 & (3.5 $\pm$ 0.8) $\cdot$ 10$^{-1}$ & 18.5 $\pm$ 1.4 & (2.3 $\pm$ 0.6) $\cdot$ 10$^{-1}$  \\ 
with natural K &   &  &  &  & &   \\ \hline

Assembled PMT (2015) & (3.8 $\pm$ 0.7) $\cdot$ 10$^{-1}$  & < 1.6 & < 32 &  (2.9 $\pm$ 0.6) $\cdot$ 10$^{-1}$  & < 1.4 & (2.2 $\pm$ 0.5) $\cdot$ 10$^{-1}$  \\ 
with $^{39}$K-enriched &   &  &  &  &  & \\ \hline 

Part-by-part sum (2015) &  (6.2 $\pm$ 1.7) $\cdot$ 10$^{-1}$  &  < 2.2  & < 1.6 $\cdot$ 10$^{2}$ & (4.8 $\pm$ 1.5) $\cdot$ 10$^{-1}$  & < 1.3   & (4.8 $\pm$ 0.6) $\cdot$ 10$^{-1}$ \\ \hline \hline

Assembled PMT (2016) &  (4.4 $\pm$ 0.6) $\cdot$ 10$^{-1}$   & < 1.4 & < 24 & (2.0 $\pm$ 0.6) $\cdot$ 10$^{-1}$  & 2.0 $\pm$ 0.5  & (1.3 $\pm$ 0.4) $\cdot$ 10$^{-1}$  \\ 
with $^{39}$K-enriched &   &  &  &  &  &  \\ \hline \hline
Part-by-part sum (2016) & (3.9 $\pm$ 1.0) $\cdot$ 10$^{-1}$   &  < 2.0 & < 1.1 $\cdot$ 10$^{2}$ & < 2.0 $\cdot$ 10$^{-1}$ & < 1.0 & (1.9 $\pm$ 0.4) $\cdot$ 10$^{-1}$    \\ \hline

   \end{tabular}
  }
  \label{tab:comparison}
\end{table}

As summarized in Table \ref{tab:comparison}, 
we achieved a large reduction of  $^{40}$K contamination by using $^{39}$K enriched potassium for
the photocathode. PMTs produced early in 2015 were made with potassium of natural isotope abundance,
leading to 18.5 $\pm$ 1.4 mBq/PMT from $^{40}$K.
Later PMT batches produced with $^{39}$K enriched potassium yielded an upper limit of 1.4 mBq/PMT
for PMTs produced in 2015 and a measurement of 2.0 $\pm$ 0.5 mBq/PMT for those produced in 2016.
Using $^{39}$K enriched potassium thus reduced $^{40}$K decays in these PMTs by almost one order of magnitude.

To study the contamination during the production process,
we compare the sum of all parts to the measurements of assembled PMTs.
Table \ref{tab:comparison} shows the sum of all parts and the assembled PMTs 
for the year 2015 and 2016 with the enriched potassium.
Among these numbers only $^{60}$Co for the sum of all parts for 2015 seems to
hint at some inconsistency: 
0.48 $\pm$ 0.06 mBq/PMT for $^{60}$Co is significantly larger than
the measurements on the assembled PMTs.
This difference is explained by the decay of $^{60}$Co. 
The components were measured in 2015, while
the assembled PMT measurements were made in 2018,  
after the installation of the ultra-low-BG HPGe detector.
After assembly, the PMTs were stored in the underground where the flux of cosmic-ray muons
is much smaller and $^{60}$Co cosmogenic production is negligible compared to that at ground level. 
Three years decrease the amount of
$^{60}$Co to 33\%, and thus the corrected amount of $^{60}$Co for the sum of components in 2018 
then becomes 
0.32 $\pm$ 0.04 mBq/PMT,
which is consistent with the measurements of PMTs assembled in 2018.
Since no significant difference between the sum of the components and the assembled PMTs
was identified for all RI components,
we concluded that there was no significant contamination introduced during the assembling process.

\subsubsection{Comparison to other PMT used for LXe detectors}
Table \ref{tab:compto10789} shows the photocathode area normalized RI summaries
for the R13111 PMT and
the R10789 PMT.
The values for the R10789 PMT are normalized using its photocathode area of 24.6 cm$^{2}$.
The R13111 PMT values were normalized with the projected area its
photocathode presents for normal incidence at its center:
a circle of 70 mm diameter with an area of 38.5 cm$^{2}$.
When compared to the R10789 PMT, $^{226}$Ra, and $^{228}$Ra were reduce by about a factor of 5,
the $^{40}$K by a factor larger than 7 and $^{60}$Co by a factor larger than 20.
Thus with the R13111 PMT we did indeed achieve a large reduction of RI. 

Table 9 compares the RIs between the R13111 PMT variants presented
in this work and the R11410 PMT used 
by XENON1T\cite{Aprile:2015pmt}, PandaX\cite{Cao:2014jsa}
and LUX\cite{Akerib:2012da}.
Almost all PMT RIs are lowest for our R13111 among these PMTs.
Only $^{226}$Ra for the 2016 R13111 PMTs
is larger than the R11410-10 (LUX) number,
though the levels are quite similar.

\begin{table}[htb]
  \caption{Results of measurements for the R13111 PMT and the R10789 PMT
    in $\mu$Bq/cm$^{2}$. The numbers are normalized
    by photocathode area.  Errors in the table are statistical only. See the text for the systematic error.}
\scalebox{1.0}{
\begin{tabular}{|c|r|r|r|r|r|}
\hline
$\mu$Bq/cm$^{2}$ & $^{226}$Ra  &  $^{228}$Ra & $^{40}$K & $^{60}$Co  \\
\hline
\hline
Assembled PMT (2015) & 9.9 $\pm$ 1.8  & 7.5 $\pm$ 1.6 & <36 &5.7 $\pm$ 1.3\\ 
Assembled PMT (2016) & 11 $\pm$ 2 & 5.2 $\pm$ 1.6 & 52 $\pm$ 13  &3.4 $\pm$ 1.0\\ 
\hline
R10789          & 49 $\pm$ 12 & <32 & (3.7 $\pm$ 0.9)$\cdot$ 10$^{2}$ &(1.1 $\pm$ 0.1)$\cdot$ 10$^{2}$\\ 
\hline

   \end{tabular}
  }
  \label{tab:compto10789}
\end{table}

\begin{table}[htb]
\caption{RI comparison between the R13111 PMT and other PMTs used by other groups. The units are $\mu$Bq/PMT.}
\scalebox{0.9}{
\begin{tabular}{|l|r|r|r|r|r|}
\hline
$\mu$Bq/PMT &  $^{226}$Ra  & $^{238}$U &  $^{228}$Ra & $^{40}$K & $^{60}$Co  \\
\hline
\hline
R13111 in 2015 & (3.8 $\pm$ 0.7)$\cdot$ 10$^{2}$ & <1.6 $\cdot$ 10$^{3}$ & (2.9 $\pm$ 0.6)$\cdot$ 10$^{2}$ & <1.4    $\cdot$ 10$^{3}$  &(2.2 $\pm$ 0.5)$\cdot$ 10$^{2}$\\ 
R13111 in 2016 & (4.4 $\pm$ 0.6)$\cdot$ 10$^{2}$& <1.4  $\cdot$ 10$^{3}$ & (2.0 $\pm$ 0.6)$\cdot$ 10$^{2}$ & (2.0 $\pm$ 0.5) $\cdot$ 10$^{3}$ &(1.3 $\pm$ 0.4)$\cdot$ 10$^{2}$\\ 
\hline
R11410-21(XENON1T)\cite{Aprile:2015pmt}& (5.2 $\pm$ 1.0)$\cdot$ 10$^{2}$& <1.3 $\cdot$ 10$^{4}$ & (3.9 $\pm$ 1.0)$\cdot$ 10$^{2}$ & (1.2 $\pm$ 0.2) $\cdot$ 10$^{4}$ &(7.4 $\pm$ 1.0) $\cdot$ 10$^{2}$\\ 
R11410-10(PandaX)\cite{Cao:2014jsa}& <7.2$\cdot$ 10$^{2}$& ---  & <8.3$\cdot$ 10$^{2}$ & (1.5 $\pm$ 0.8) $\cdot$ 10$^{4}$ & (3.4 $\pm$ 0.4) $\cdot$ 10$^{3}$ \\ 
R11410-10(LUX)\cite{Akerib:2012da}  & <4.0$\cdot$ 10$^{2}$ & <6.0 $\cdot$ 10$^{3}$ & <3.0 $\cdot$ 10$^{2}$ & <8.3 $\cdot$ 10$^{3}$     & (2.0 $\pm$ 0.2) $\cdot$ 10$^{3}$ \\ 
\hline

   \end{tabular}
  }
  \label{tab:compto11410}
\end{table}

\section{Afterpulse study related to the Xe leak problem}
\label{sec:Afterpulse}

For the R11410, which has many structures in common with the R13111,
Xe afterpulses caused
by Xe leaking into the PMTs were previously reported by XENON1T and PandaX \cite{Barrow:PMT, Li:PMT}.
Such Xe leaks are a serious problem for LXe detectors, since afterpulses interfere with
and confuse the dark matter or other physics signals; 
a large enough leak ultimately makes it impossible to operate such a PMT.
Therefore, we examined the Xe leak for the R13111 PMT and observed a similar
leak as reported for R11410.
We carried out a detailed study of the Xe afterpulse rate to solve this problem.

\subsection{Identification of the leaking component}
We observed no Xe leak in 642 R10789 PMTs in XMASS-I,
while leaks were observed in both the R13111 and the R11410 PMT.
We compared the structure of these three PMTs and narrowed down the possible causes. 
First, we compared the R10789 with the R13111.
The components modified from the R10789 to the R13111 could be the cause.
Among the modifications, there were two candidates which might have caused the leak.
One is the seal between the window and the body, the other are the glass beads at the stem.  
The R13111 PMT uses 5N aluminum for the seal while the R10789 PMT uses 98.5\% aluminum.
The glass beads are the electrical insulation at the feedthroughs at the stem and
changed to ones made from high-purity ingredients in the R13111 PMT.
Next we compared the R13111 with the R11410.
We hypothesized that the cause of the leak would be common in these two PMTs.
From this second comparison,
we narrowed down the possible causes to the 5N aluminum seal which is used in both PMTs.
The high purity glass beads are not used in the R11410 PMT.
Hence, the surfaces of the seals in the assembled R10789 and the R13111 PMT were examined
in detail and it was found that the seal surface of the R13111 PMT had many tiny scratches
compared to that of the R10789 PMT. 
Consequently, we investigated the following two hypothetical causes of the leak;
one is the usage of high-purity 5N aluminum itself and the other is
a deterioration of the seal surface condition caused by the exchange itself
not by the purity change.

To check this, we prepared two types of the R11410 PMT. 
One uses lower purity 4N aluminum for the seal, 
the other uses the usual 5N aluminum, however handling of the seal during the PMT
assembling process was revised to pay special attention to the surface of the seal.

\subsection{Setup of the study}
\begin{figure}\centering
  \includegraphics[width=7cm]{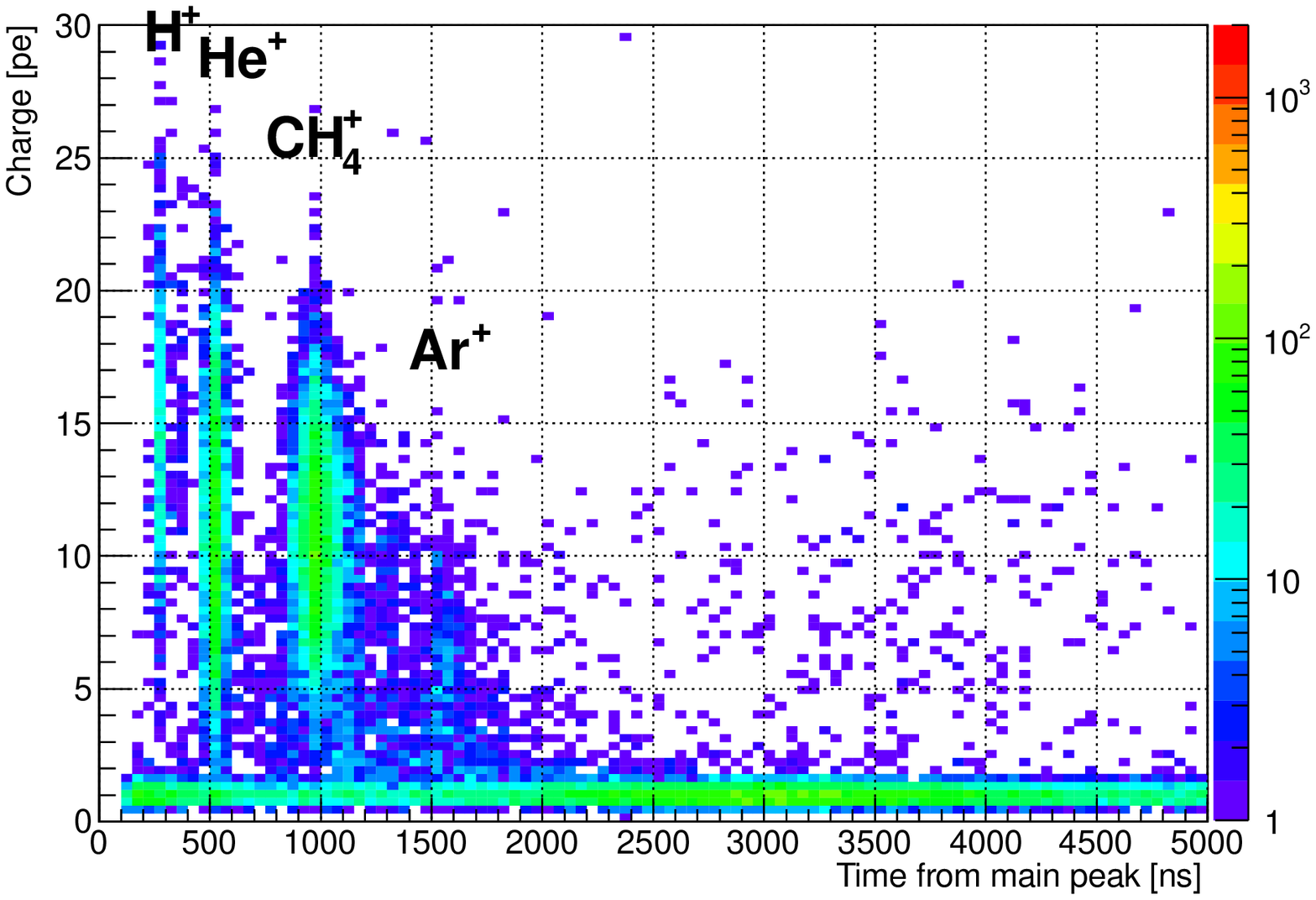}
    \includegraphics[width=7cm]{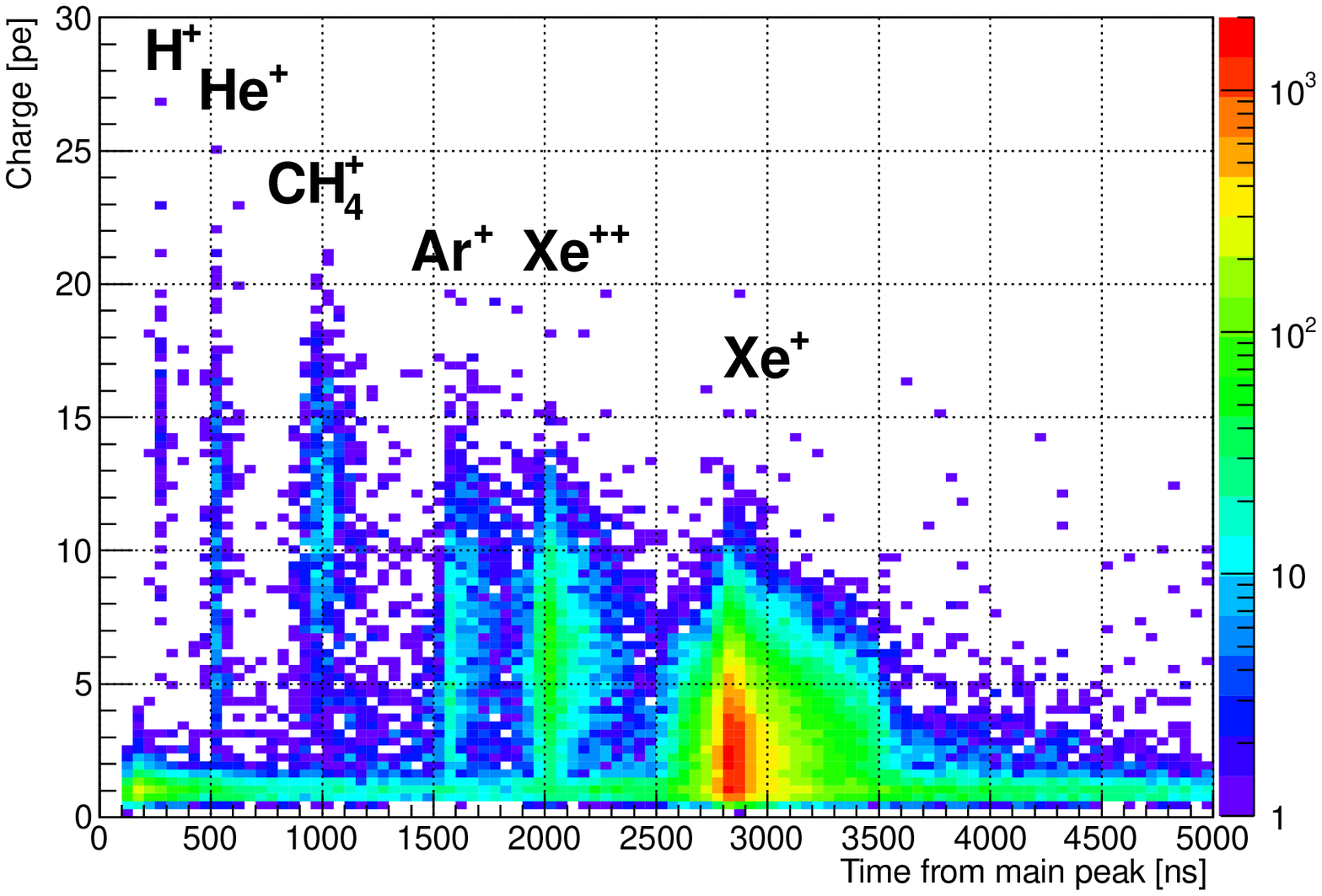}
    \caption{Afterpulse charge versus time delay from a light source signal for a normal PMT (left)
      and a PMT with Xe leak(right)
      at room temperature.
      Afterpulses from different ions of the residual gas in the PMTs vacuum
      were observed above the one photoelectron band 
      at their respective expected timing.}
  \label{fig:afterpulse}
\end{figure}

To examine the afterpulse in a Xe atmosphere, we prepared a special chamber 
inside of which the PMTs could be cooled down to the LXe temperature of about -100 degree Celsius
in a gaseous Xe atmosphere.
By applying multiple temperature cycles of cooling and warming, 
we examined the afterpulses caused by ions of the residual gas.
We applied at least three cooling cycles to each PMT. 
A light source (LED or laser diode) was adjusted to a level
in which the PMT mostly observes just one photoelectron per flash of the respective light source.
Afterpulses which follow the one photoelectron light source signal were recorded with FADCs
and examined.
Figure \ref{fig:afterpulse} shows the afterpulse charge against
the time delay from the light source signal 
for a PMT without the Xe leak (left hand side)
and one with a Xe leak (right hand side).
All PMTs were operated at a gain of 1 $\times$ 10$^{7}$ in the study.

\subsection{Results of the study}
The number of PMTs that developed leaks after the temperature cycles in the study is
summarized in Table \ref{tab:xeleak}.
The Xe leaking PMTs were separated into two categories by their respective afterpulse rate,
one is ``large leak'' and the other is ``small leak''.
The afterpulse rate is the probability of afterpulse observation after one light source signal.
We found that these two categories have different properties. 
For the ``large leak'' category the Xe afterpulse rate
increased with every cooling cycle.
The Xe afterpulse rate of all the large leak PMTs
exceeded 0.1\% after the third cooling cycle.   
For the ``small leak” category,
the Xe afterpulse rates are below 0.1\% for all the small leak PMT even
after the third cooling cycle.
There was no increase in the afterpulse rate by the second and the third cooling cycle.
We confirmed this property of the small leak category by applying additional cooling cycles to
PMTs in this category. After applying 
seven additional cooling cycles to 10 small leak R13111 and one R11410 PMTs,
we observed no change in their afterpulse rate.

Thirty of the 58 examined R13111 PMTs fabricated with 5N aluminum were also measured in
liquid Xe rather than only cold gaseous Xe.
Seven large and 3 small leaks among these 30 PMTs show no significant difference from 9 large
and 7 small leaks observed for the other 28 PMTs tested only in gaseous Xe.
The R11410 PMTs were only examined in gaseous Xe.

Out of 58 R13111 PMTs 16 developed large leaks and 10 small leaks during the cooling cycles.
When compared to the numbers for R11410 with 5N aluminum reported as 
7 large leaks and one small leak
among 44 PMTs \cite{ Aprile:2015pmt},
the fraction of Xe leaks for R13111 appears to be larger than that of R11410.
The fraction of large leaks is 28 $\pm$ 6\% for R13111 and 16 $\pm$ 6\% for R11410,
both with 5N Al.
For the 4N aluminum R11410 PMT, 
we observed no large leak and one small leak.
The probability of a large leak is less than 10\% with 90\% CL, 
assuming the observed fraction of the leak represents the probability of a leak.
We also assumed no probability difference among PMT production batches
if we keep the same assembling process.
Though the significance is not large, the use of lower purity aluminum for the seal
reduce the leak probability.
The same improvement can be seen for the PMTs with 5N aluminum
that were prepared with special attention to the seal 
with an ever greater level of confidence.
For these PMTs, 
neither large nor small leaks were observed in 28 PMTs.
Consequently, the probability of large leak is less than 8\%.

We studied and confirmed more leaks in PMTs which used 5N aluminum for their seals than
the ones which used 4N aluminum.
This suggests 
that the exchange of the sealing material to 5N aluminum is a cause of the Xe leaks.
We also observed that it is possible to suppress leakage when using 5N aluminum 
by applying appropriate surface care.
Since the sealing structure is common in the R11410 and the R13111,
this result obtained with the R11410 PMT should also apply for the R13111 PMT.

\begin{table}[htb]
\caption{Results of the Xe leak test for the R13111 PMTs and the R11410 PMTs}
\scalebox{1.0}{
\begin{tabular}{|c|l|r|r|r|r|}
\hline
PMT  & Sealing type
     & Tested number & Large leaking & Small leaking & Large leak\\
     & &        &           PMT &           PMT & probability\\ 
     & &        & Xe afterpulse & Xe afterpulse &  \\
     & &        &    \(>\)0.1\% &  \(<\)0.1\%   & \\
\hline
R13111 & 5N Al & 58 & 16 & 10 & 28 $\pm$ 6\%\\
\hline
R11410 & 5N Al \cite{Aprile:2015pmt} & 44 & 7 & 1 & 16 $\pm$ 6\%\\
\hline
\hline
R11410 & 4N Al & 23 & 0  & 1& \(<\)10\%\\
\hline
R11410 &5N Al with & 28 & 0 & 0  & \(<\)8\%\\
       & special handling  &    &    &  &\\
\hline

   \end{tabular}
  }
  \label{tab:xeleak}
\end{table}

\section{Conclusion}
\label{sec:Conclusion}
We developed a new three-inch low-BG PMT, the Hamamatsu R13111 PMT, 
which can operate in LXe.

The R13111 has a convex photocathode
that gives it a large angular acceptance and improved 
timing resolution.
For all tested points on the photocathode,
relative efficiency multiplied by the quantum efficiency is higher than 70\%.
The TTS of this new PMT is 2.11 $^{+0.31}_{-0.18}$ nsec, a large improvement over
the 6.87 $\pm$ 0.31 nsec TTS for the R10789.
This helps towards more effective pulse shape discrimination.

RIs in the R13111 PMTs manufactured in 2016 are 
0.44 $\pm$ 0.06 mBq of $^{226}$Ra, $^{238}$U less than 1.4 mBq,
0.20 $\pm$ 0.06 mBq of $^{228}$Ra, 2.0 $\pm$ 0.5 mBq of $^{40}$K and 0.13 $\pm$ 0.04 mBq of $^{60}$Co.
The reduction from the R10789 values is large: factors of 5, 6, 7 and 30
for $^{226}$Ra, $^{228}$Ra, $^{40}$K, and $^{60}$Co, respectively.
These numbers are the world's best when compared to other currently available PMTs
for use in LXe dark matter search experiments.

For long-term stable operation in LXe, we studied the Xe leak problem which was reported
by several groups.
We identified that the material and treatment of the aluminum seal between the PMT's
entrance window and the body causes the Xe leaks.
When the seal was changed to high-purity 5N aluminum for the R13111 and the R11410 PMT from
98.5\% purity aluminum used for the R10789 PMT, leaks started occurring.
From our study, we observed that the problem is solved by handling the 5N aluminum
seal with sufficient care for the seal's surface during the assembly process.

With its low radioactivity, excellent timing resolution,
and large angular acceptance 
the new R13111 PMT
can show highly good performance 
in future low BG LXe detectors.

\section{Acknowledgments}
We thank Hamamatsu Photonics K. K. for the cooperation in producing the low BG PMTs.
We gratefully acknowledge the cooperation of Kamioka Mining and Smelting Company.
We also would like to thank OKAMOTO GLASS Co. Ltd. for their help in the glass
synthesis.
This work was supported by the Japanese Ministry of Education, Culture, Sports,
Science and
Technology, Grant-in-Aid for Scientific Research, ICRR Joint-Usage,
JSPS KAKENHI Grant Number,
19GS0204, 26104004, 19H05805,  
and partially
by the National Research Foundation of Korea Grant (NRF-2011-220-C00006) 
and Institute for Basic Science (IBS-R017-G1-2018-a00).


\begin{thebibliography}{99}

\bibitem{Abe:2013tc}
  K.~Abe {\it et al.} [XMASS Collaboration],
  ``XMASS detector,''
  Nucl.\ Instrum.\ Meth.\ A {\bf 716} (2013) 78
  [arXiv:1301.2815 [physics.ins-det]].

\bibitem{Akerib:2012ys}
  D.~S.~Akerib {\it et al.} [LUX Collaboration],
  ``The Large Underground Xenon (LUX) Experiment,''
  Nucl.\ Instrum.\ Meth.\ A {\bf 704} (2013) 111
  [arXiv:1211.3788 [physics.ins-det]].

\bibitem{Cao:2014jsa}
  X.~Cao {\it et al.} [PandaX Collaboration],
  ``PandaX: A Liquid Xenon Dark Matter Experiment at CJPL,''
  Sci.\ China Phys.\ Mech.\ Astron.\  {\bf 57} (2014) 1476
  [arXiv:1405.2882 [physics.ins-det]].

\bibitem{Aprile:2017aty}
  E.~Aprile {\it et al.} [XENON Collaboration],
  ``The XENON1T Dark Matter Experiment,''
  Eur.\ Phys.\ J.\ C {\bf 77} (2017) no.12,  881
  [arXiv:1708.07051 [astro-ph.IM]].
  
\bibitem{Suzuki:2018xek} 
  T.~Suzuki {\it et al.} [XMASS Collaboration],
  ``Search for WIMP-$^{129}$Xe inelastic scattering with particle identification in XMASS-I,''
  Astropart.\ Phys.\  {\bf 110}, 1 (2019)
  [arXiv:1809.05358 [astro-ph.CO]].
 
\bibitem{Kobayashi:2018jky} 
  M.~Kobayashi {\it et al.} [XMASS Collaboration],
  ``Search for sub-GeV dark matter by annual modulation using XMASS-I detector,''
  Phys.\ Lett.\ B {\bf 795}, 308 (2019)
  arXiv:1808.06177 [astro-ph.CO].
 
\bibitem{Abe:2018owy} 
  K.~Abe {\it et al.} [XMASS Collaboration],
  ``Search for dark matter in the form of hidden photons and axion-like particles in the XMASS detector,''
  Phys.\ Lett.\ B {\bf 787}, 153 (2018)
  doi:10.1016/j.physletb.2018.10.050
  [arXiv:1807.08516 [astro-ph.CO]].
  
\bibitem{XMASS:2018bid} 
  K.~Abe {\it et al.} [XMASS Collaboration],
  ``A direct dark matter search in XMASS-I,''
  Phys.\ Lett.\ B {\bf 789}, 45 (2019)
  [arXiv:1804.02180 [astro-ph.CO]].

\bibitem{Abe:2018mxq} 
  K.~Abe {\it et al.} [XMASS Collaboration],
  ``Direct dark matter search by annual modulation with 2.7 years of XMASS-I data,''
  Phys.\ Rev.\ D {\bf 97}, no. 10, 102006 (2018)
  [arXiv:1801.10096 [astro-ph.CO]].
 
\bibitem{Abe:2012az} 
  K.~Abe {\it et al.},
  ``Light WIMP search in XMASS,''
  Phys.\ Lett.\ B {\bf 719}, 78 (2013)
  [arXiv:1211.5404 [astro-ph.CO]].
  
\bibitem{Abe:2018biz} 
  K.~Abe {\it et al.} [XMASS Collaboration],
  ``Development of low radioactivity photomultiplier tubes for the XMASS-I detector,''
  Nucl.\ Instrum.\ Meth.\ A {\bf 922}, 171 (2019)
  [arXiv:1808.03617 [physics.ins-det]].

\bibitem{Mount:2017qzi} 
  B.~J.~Mount {\it et al.},
  ``LUX-ZEPLIN (LZ) Technical Design Report,''
  arXiv:1703.09144 [physics.ins-det].
  
  \bibitem{Barrow:PMT}
  P.~Barrow {\it et al.},
  ``Qualification Tests of the R11410-21 Photomultiplier Tubes for the XENON1T Detector'',
  JINST \  {\bf 12} (2017) P01024
  [arXiv:1609.01654v2 [astro-ph.IM]]

  \bibitem{Li:PMT} 
  S.~Li {\it et al.} 
  ``Performance of photosensors in the PandaX-I experiment'',
  JINST \  {\bf 11} (2016) T02005
  [arXiv:1511.06223v3 [physycs.ins-det] ]
  
  \bibitem{Aprile:2015pmt}
  E.~Aprile {\it et al.} [XENON Collaboration],
  ``Lowering the radioactivity of the photomultiplier tubes for the XENON1T dark matter experiment,''
  Eur.\ Phys.\ J.\ C {\bf 75} (2015)  546
  [arXiv:1503.07698 [astro-ph.IM]].

  \bibitem{Agostinelli:2002hh} 
  S.~Agostinelli {\it et al.} [GEANT4 Collaboration],
  ``GEANT4: A Simulation toolkit,''
  Nucl.\ Instrum.\ Meth.\ A {\bf 506} (2003) 250.

\bibitem{MS:nanos}
  Eurofins EAG Materials Science Tokyo Corporation,
  http://www.nanoscience.co.jp/
  
\bibitem{MS:toray}
  Toray Research Center,
  https://www.toray-research.co.jp/
  
  \bibitem{Akerib:2012da} 
  D.~S.~Akerib {\it et al.},
  ``An Ultra-Low Background PMT for Liquid Xenon Detectors,''
  Nucl.\ Instrum.\ Meth.\ A {\bf 703}, 1 (2013)
  [arXiv:1205.2272 [physics.ins-det]].


\bibitem{Xu:2016pha} 
  B.~Xu [XMASS Collaboration],
  ``XMASS 1.5: The next step in Kamioka, Japan,''
  J.\ Phys.\ Conf.\ Ser.\  {\bf 718}, no. 4, 042064 (2016).


\end{thebibliography}
\end{document}